\newcommand{\Ntotal}  {120\,}
\newcommand{\intLdt}  {10.4\,}
\newcommand{\intLdtfull}  {10.36\,}

\newcommand{\dLtot}   {0.06\,}
\newcommand{\rroots}  {172.12\,}  
\newcommand{\droots}  {0.06\,}

\newcommand{\LepII}{\mbox{LEP2}}
\newcommand{\LepI}{\mbox{LEP1}}

\newcommand{\Com}{centre-of-mass}
\newcommand{\SM}{Standard Model}
\newcommand{\MC}{Monte Carlo}
\newcommand{\tgc}{{\small TGC}}

\newcommand{\GeV}{\mbox{$\mathrm{GeV}$}}

\newcommand{\Ipb}{\mbox{pb$^{-1}$}}
\newcommand{\beq}{\begin{equation}}
\newcommand{\eeq}{\end{equation}}
\newcommand{\bea}{\begin{eqnarray}}
\newcommand{\eea}{\end{eqnarray}}
\newcommand{\ra}{\rightarrow}

\newcommand{\bm}{\boldmath}
\renewcommand{\deg}{^\circ}
\def\gappeq{\mathrel{ \rlap{\raise.5ex\hbox{$>$}}
                      {\lower.5ex\hbox{$\sim$}}  } }
\def\lappeq{\mathrel{ \rlap{\raise.5ex\hbox{$<$}}
                      {\lower.5ex\hbox{$\sim$}}  } }

\newcommand{\etal}{{\it et al.}} 
\newcommand{\PLB}[3]  {Phys.\ Lett.\ {\bf B#1} (#2) #3}
\newcommand{\ZPC}[3]  {Z.\ Phys.\ {\bf C#1} (#2) #3}
\newcommand{\ZFPC}    {Z.\ Phys.\ C}
\newcommand{\NIMA}[3] {Nucl.\ Instr.\ Meth.\ {\bf A#1} (#2) #3}

\newcommand{\PRL}[3]  {Phys.\ Rev.\ Lett.\ \textbf{#1} (#2) #3}
\newcommand{\PRD}[3]  {Phys.\ Rev.\ {\bf D#1} (#2) #3}
\newcommand{\NPB}[3]  {Nucl.\ Phys.\ {\bf B#1} (#2) #3}

\newcommand{\CPC}[3]  {Comp.\ Phys.\ Comm.\ {\bf #1} (#2) #3}
\def\opalakrawy{The OPAL Collaboration, M.Z.\ Akrawy \etal}

\newcommand{\Mw}{\mbox{$M_{\mathrm{W}}$}}
\newcommand{\Gw}{\mbox{$\Gamma_{\mathrm{W}}$}}

\newcommand{\cwsq}{\mbox{$\cos^2\theta_w$}}
\newcommand{\twsq}{\mbox{$\tan^2\theta_w$}}
\newcommand{\epem}{\mbox{$\mathrm{e^+e^-}$}}

\newcommand{\Zz}{\mbox{${\mathrm{Z}^0}$}}
\newcommand{\WW}{\mbox{$\mathrm{W^+W^-}$}}
\newcommand{\Wm}{\mbox{$\mathrm{W^-}$}}

\newcommand{\qq}{\mbox{$\mathrm{q\overline{q}}$}}
\newcommand{\Wqq}{\mbox{$\mathrm{q\overline{q} }$}}
\newcommand{\Wqqp}{\mbox{$\mathrm{q\overline{q} }$}}

\newcommand{\lnu}{\mbox{$\ell\overline{\nu}_{\ell}$}}
\newcommand{\lnubar}{\mbox{$\overline{\ell}^\prime\nu_{\ell^\prime}$}}

\newcommand{\enu}{\mbox{$\mathrm{e\overline{\nu}_{e}}$}}
\newcommand{\mnu}{\mbox{$\mu\overline{\nu}_{\mu}$}}
\newcommand{\tnu}{\mbox{$\tau\overline{\nu}_{\tau}$}}
\newcommand{\WWqqqq}{\mbox{\WW$\rightarrow$ \Wqq\Wqqp}}

\newcommand{\WWqqen}{\mbox{\WW$\rightarrow$ \Wqq\enu}}
\newcommand{\WWqqmn}{\mbox{\WW$\rightarrow$ \Wqq\mnu}}
\newcommand{\WWqqtn}{\mbox{\WW$\rightarrow$ \Wqq\tnu}}
\newcommand{\WWlnln}{\mbox{\WW$\rightarrow$ \lnu\lnubar}}

\newcommand{\Zqq}{\mbox{$\Zz\rightarrow\qq$}}

\newcommand{\ZGqq}{\mbox{$\Zz/\gamma\rightarrow\qq$}}

\newcommand{\roots}{\mbox{$\sqrt{s}$}}

\newcommand{\Zgamma}{\mbox{$\Zz/\gamma$}}

\newcommand{\Cthw}{ \mbox{$\cos\theta_{\mathrm{W}} $}}

\newcommand{\Cthstl}{\mbox{ $\cos \theta_\ell^{*}$}}
\newcommand{\Phistl}{\mbox{ $\phi_\ell^{*}$}}

\newcommand{\Lnln}{\lnu\lnubar}
\newcommand{\Qqln}{\Wqq\lnu}
\newcommand{\Qqen}{\Wqq\enu}
\newcommand{\Qqmn}{\Wqq\mnu}
\newcommand{\Qqtn}{\Wqq\tnu}
\newcommand{\Qqqq}{\Wqq\Wqqp}

\newcommand{\gz}{\mbox{$g_1^{\mathrm{z}}$}}
\newcommand{\kg}{\mbox{$\kappa_\gamma$}}
\newcommand{\kz}{\mbox{$\kappa_{\mathrm{z}}$}}
\renewcommand{\lg}{\mbox{$\lambda_\gamma$}}
\newcommand{\lz}{\mbox{$\lambda_{\mathrm{z}}$}}
\newcommand{\dgz}{\mbox{$\Delta g_1^{\mathrm{z}}$}}
\newcommand{\dk}{\mbox{$\Delta \kappa$}}
\newcommand{\dkg}{\mbox{$\Delta \kappa_\gamma$}}
\newcommand{\dkH}{\mbox{$\Delta \kappa_\gamma^{\tiny(HISZ)}$}}
\newcommand{\dkz}{\mbox{$\Delta \kappa_{\mathrm{z}}$}}

\newcommand{\abf}{\mbox{$\alpha_{B\phi}$}}
\newcommand{\awf}{\mbox{$\alpha_{W\phi}$}}
\newcommand{\aw}{\mbox{$\alpha_W$}}

\newcommand{\mln}{M_{\ell\nu}}

\newcommand{\LL}{\log L}

\newcommand{\resabf} {$0.35^{+1.29}_{-1.07}\pm0.38$} 
\newcommand{\resawf} {$0.00^{+0.30}_{-0.28}\pm0.11$}
\newcommand{\resaw}  {$0.18^{+0.49}_{-0.47}\pm0.23$}
\newcommand{\resdgz} {$-0.03^{+0.40}_{-0.37}\pm0.14$}
\newcommand{\resdkH} {$0.03^{+0.55}_{-0.51}\pm0.20$}
\newcommand{\resdk}  {$0.03^{+0.49}_{-0.46}\pm0.21$}
\newcommand{\resawfc} {$-0.08^{+0.28}_{-0.25}\pm0.10$}

\documentstyle[epsfig,amsbsy,a4p,12pt]{article}

\begin{document}
\bibliographystyle{plain}
\begin{titlepage}
 \begin{center}{\large   EUROPEAN LABORATORY FOR PARTICLE PHYSICS
 }\end{center}\bigskip
\begin{flushright}
  CERN-PPE/97-125 \\ 
16 September 1997 \\
\end{flushright}
\bigskip\bigskip\bigskip\bigskip\bigskip
\begin{center}
 {\LARGE\bf \boldmath 
Measurement of triple gauge boson couplings from 
\WW\, production at $\roots=172~\GeV$}
\end{center}
\bigskip\bigskip
\begin{center}{\Large The OPAL Collaboration
}\end{center}\bigskip
\bigskip
\bigskip\begin{center}{\large  Abstract}\end{center}
We present 
measurements of triple gauge boson coupling parameters 
using data recorded by the OPAL detector at \LepII\,
at a centre-of-mass energy of 172 \GeV.
A total of \Ntotal W-pair candidates has been selected 
in the \Qqqq, \Qqln\, and \Lnln\, decay channels, for an integrated
luminosity of \intLdt\Ipb. 
We use these data to determine
several different anomalous coupling parameters
using the measured cross-section 
and the distributions of kinematic variables.
We measure 
\abf=\resabf ,
\awf=\resawf ,
\aw=\resaw ,
\dgz=\resdgz ,
\dkH=\resdkH , and
\dk=\resdk .
Combining the $\awf$ result with our previous
result obtained from the 161 \GeV\, data sample 
we measure
\awf=\resawfc .
All of these measurements are consistent with the \SM .
\bigskip\bigskip\bigskip\bigskip
\bigskip\bigskip
\begin{center}{\large
(To be submitted to Zeitschrift f\"{u}r Physik C.)
}\end{center}



\end{titlepage}
\begin{center}{\Large        The OPAL Collaboration
}\end{center}\bigskip
\begin{center}{
K.\thinspace Ackerstaff$^{  8}$,
G.\thinspace Alexander$^{ 23}$,
J.\thinspace Allison$^{ 16}$,
N.\thinspace Altekamp$^{  5}$,
K.J.\thinspace Anderson$^{  9}$,
S.\thinspace Anderson$^{ 12}$,
S.\thinspace Arcelli$^{  2}$,
S.\thinspace Asai$^{ 24}$,
D.\thinspace Axen$^{ 29}$,
G.\thinspace Azuelos$^{ 18,  a}$
% ...
}\end{center}\bigskip
\bigskip
% ...
$^{  2}$Dipartimento di Fisica dell' Universit\`a di Bologna and INFN,
I-40126 Bologna, Italy
\newline
% ...
$^{  5}$Cavendish Laboratory, Cambridge CB3 0HE, UK
\newline
% ...
$^{  8}$CERN, European Organisation for Particle Physics,
CH-1211 Geneva 23, Switzerland
\newline
$^{  9}$Enrico Fermi Institute and Department of Physics,
University of Chicago, Chicago IL 60637, USA
\newline
% ...
$^{ 12}$Indiana University, Department of Physics,
Swain Hall West 117, Bloomington IN 47405, USA
\newline
% ...
$^{ 16}$Department of Physics, Schuster Laboratory, The University,
Manchester M13 9PL, UK
\newline
% ...
$^{ 18}$Laboratoire de Physique Nucl\'eaire, Universit\'e de Montr\'eal,
Montr\'eal, Quebec H3C 3J7, Canada
\newline
% ...
$^{ 23}$Department of Physics and Astronomy, Tel Aviv University,
Tel Aviv 69978, Israel
\newline
$^{ 24}$International Centre for Elementary Particle Physics and
Department of Physics, University of Tokyo, Tokyo 113, and
Kobe University, Kobe 657, Japan
\newline
% ...
$^{ 29}$University of British Columbia, Department of Physics,
Vancouver BC V6T 1Z1, Canada
\newline
% ...
$^{  a}$ and at TRIUMF, Vancouver, Canada V6T 2A3
\newline
% ...
\newline
\pagebreak
\section{Introduction}
 \label{sec-introduction}

During the initial year of operation of \LepII, centre-of-mass energies 
have been attained which allow the production of \WW\, boson pairs 
for the first time in \epem\, collisions. 
A total integrated luminosity of 9.9~\Ipb\, was recorded at a \Com\
energy of 161 \GeV\, and \intLdt\Ipb\, at 172 \GeV.

The \WW\, production process involves the 
triple gauge boson vertices between the \WW\, and the \Zz\, or photon.
The measurement of these triple gauge boson couplings (\tgc s) and the
search for possible anomalous values is one of the principal 
physics  goals at \LepII. 
Previous direct measurements have been presented by 
CDF~\cite{CDF} and D0~\cite{D0}, 
and by the LEP experiments
using the 161 \GeV ~\cite{tgc161-analysis,OTHERLEP161-tgc}
and 172 \GeV~\cite{OTHERLEP172-tgc} data.
In this paper we present an analysis
of the  172 \GeV\, data sample to make measurements of several
anomalous coupling parameters. 
We also combine these results
with the measurement obtained from our 161 \GeV\, 
data sample~\cite{tgc161-analysis}. 

The observable effects of possible deviations from the \SM\, 
have been studied extensively~\cite{LEP2YR}.
Anomalous \tgc s can affect both the total
production cross-section and the shape of the differential
cross-section as a function of the W$^-$ production angle.
Additionally, the relative contributions of each helicity state of 
the W bosons are 
changed, which in turn affects the distributions of their 
decay products.
In this analysis we make use of both production rate 
and angular distributions measured using 
events where the W boson pair decays to the \Qqln\, 
channel (hereafter \Qqln\, also refers to the charge conjugate state). 
We also use the measured production rate for events
where the W boson pair decays to the \Qqqq\, and \Lnln\, channels.

\vspace{8mm}
{\bf Anomalous coupling parameters}
\vspace{2mm}

The most general Lorentz invariant 
Lagrangian~\cite{LEP2YR,HAGIWARA,BILENKY,GAEMERS} 
which describes the triple gauge boson interaction
has fourteen independent terms, seven describing  the WW$\gamma$ 
vertex and seven describing the WWZ vertex.  
This parameter space is very large, and it is not currently possible
to measure all fourteen couplings independently. 
Assuming electromagnetic gauge invariance  and C and P conservation
the number of parameters reduces to five,  which can be taken
as $\gz, \kz, \kg, \lz$ and $\lg$~\cite{LEP2YR,HAGIWARA}. 
In the \SM\, $\gz = \kz = \kg = 1$ and
$\lz = \lg = 0$.

Different sets of parameters have also been proposed which are
motivated by SU(2)$\times$U(1) gauge
invariance and constraints arising from precise measurements at \LepI.
One such set proposes three independent linear combinations 
of the couplings \cite{LEP2YR,BILENKY} 
which are not tightly constrained~\cite{DERUJULA,HISZ} 
by existing \LepI\, data.
These are:
\begin{eqnarray}
\abf & \equiv & \dkg - \dgz \cwsq \\
\awf & \equiv & \dgz \cwsq    \\
\aw  & \equiv & \lg .
\end{eqnarray}                               
with the constraints that
$\dkz  =  -\dkg \twsq +  \dgz$  and
$\lz  =  \lg$. 
The $\Delta$ indicates the deviation of the respective quantity from its
\SM\, value and $\theta_w$ is the weak mixing angle. Each of the $\alpha$
parameters has the value zero in the \SM.
A similar approach, HISZ~\cite{HISZ}, reduces this set to 
two parameters with the extra constraint $\abf=\awf$.
This extra constraint is equivalent to $\dgz = \dkg / (2 \cwsq)$,
and $\dkg$ and $\lg\equiv\aw$ are normally used as the variable
parameters.

In this analysis, we present measurements of the following parameters:

\begin{itemize}

\item $\abf$, $\awf$ and $\aw$,

\item the two and three dimensional correlations of $\abf$, $\awf$ and $\aw$,

\item $\dgz$
and the two dimensional correlation of $\dgz$ and $\aw$,  

\item $\dkg$ assuming the HISZ constraints,
and the two dimensional correlation of $\dkg$ and $\aw$,

\item $\dk$ assuming $\dkz = \dkg$, 
and the two dimensional correlation of $\dk$ and $\aw$.

\end{itemize}
 
In each case, all parameters not mentioned explicitly are set to
their \SM\, values. The measurements of \dkg\, and \dk\, are included
to facilitate comparisons with results from the CDF and D0 collaborations.

\section{Data selection and Reconstruction}
\label{sec-selection}

The data were recorded at an 
average \Com\, energy~\cite{LEPENERGY172} 
of $\sqrt{s}=\rroots\pm\droots$~\GeV\, using the OPAL detector, 
which is fully described elsewhere
\cite{OPAL,SW}.
A total integrated luminosity of $\intLdtfull\pm\dLtot\Ipb$ 
was recorded at this energy.
Events are selected corresponding to 
different W pair decay combinations. These are: 
(i) $\Qqqq$ events where both W bosons decay to a quark-antiquark 
final state,
(ii) $\Qqln$  events where one W decays to a 
quark-antiquark final state, and the other W decays to an 
electron, muon or tau, plus a neutrino and
(iii) $\Lnln$ events where both W bosons decay leptonically.

The selection procedures which have been developed for these channels
are based upon a multivariate likelihood method. These procedures
are described fully in~\cite{mass172-analysis}
and the results presented there are summarised in table~\ref{tab-summary}.
In this analysis we use all
of the efficiencies, background and systematic error evaluations
exactly as presented in~\cite{mass172-analysis} 
to determine the constraint
upon anomalous couplings arising from the observed event rate.

We also constrain anomalous couplings 
using the production
and decay angular distributions in the \Qqln\, channels.
We choose to use only these channels as they are the most straightforward 
to reconstruct and there is no ambiguity in assigning decay fermion 
pairs to each W,  nor in determining the charge of each W through
the decay lepton charge. 
For this part of the analysis we make additional requirements
to those described in~\cite{mass172-analysis} in order to ensure that the
charged lepton is well identified and to further reduce
background contamination. 
In the following we describe the reconstruction of \Qqln\, events
used for the angular distribution analysis and
the extra selection requirements and the corresponding 
reevaluation of the backgrounds.  

All \MC\, samples which are used
are generated assuming the current central value of the world 
average W boson mass of $\Mw=80.33 \pm 0.15$~\GeV\, \cite{MWPDG}.  
The total \WW\, cross section is assumed to be 12.4 pb 
as obtained from the GENTLE~\cite{GENTLE} program.
All \MC\, samples used are passed through the full OPAL 
simulation program~\cite{GOPAL} and then subjected to the same 
reconstruction 
procedures as applied to the data.

\subsection{\bm \Qqen\, and \Qqmn\, events} 
\label{sec-qqln}

The \Qqen\, and \Qqmn\, events are characterised by two 
well-separated hadronic jets, a high momentum charged lepton and 
missing momentum from the unobserved neutrino. 
The selections described in~\cite{mass172-analysis} 
result in a single track being identified
as the most likely lepton candidate. 
The electron momentum is constructed from the 
direction measured by the tracking detectors and the energy measured in 
the electromagnetic calorimeters.
In the case of muons, the momentum measured using the 
tracking detectors is used.
The remaining tracks and 
calorimeter clusters in the event are grouped into two jets
using the $k_\perp$ algorithm~\cite{DURHAM}. 
The total energy and momentum of each of the jets are 
calculated using both tracks and electromagnetic clusters
using the method of~\cite{GCE}. 
The main residual background is from misidentified \Qqtn\, events. 
We reduce this further by demanding 
$\mln > 50$ \GeV, where $\mln$ is the invariant
mass constructed from the lepton
four-momentum and the missing three-momentum 
assuming a zero mass. 
This extra requirement results in efficiencies of 
85\% and 87\% for the electron and muon channels respectively,
which is a loss of 0.5\% in each channel with respect to 
reference~\cite{mass172-analysis}.
The expected number of signal events and the number of events 
selected from the data 
are shown in table~\ref{tab-summary}. The values in this table
correspond to an integrated luminosity of 
$\intLdtfull\pm\dLtot$\Ipb\, at $\rroots\pm0.06$~GeV.
Table~\ref{tabtgc-eff-back} shows the signal and background cross sections
used in the analysis of the angular distributions.
The expected number of signal events is evaluated using 
PYTHIA~\cite{PYTHIA} and is
normalised such that the total cross section agrees with GENTLE.

\subsection{\bm \Qqtn\, events} 
\label{sec-qqtn}

The signature of \Qqtn\, events is less well defined than
that of the electron or muon channels due to the presence of one or more 
additional neutrinos, and the absence of a high momentum
lepton. 
The primary selection for this channel described 
in~\cite{mass172-analysis} results in the identification
of the most likely one or three track tau decay candidate
classified as an electron, muon, one-prong hadronic or three-prong
hadronic decay.
The remaining tracks and 
calorimeter clusters in the event are grouped into two jets
as described above.

The ensuing analysis of this channel relies upon the
correct tau decay products being identified in order that they
give a reasonable approximation to the original tau flight direction,
and that the direction of the hadronic W system can be reliably measured.
We therefore make additional requirements.
In the electron and muon decay channels we require the
lepton momentum to be greater than 5 \GeV.
We demand that the track is separated from other activity in the detector
by requiring that the angle between the
lepton direction and both jet axes be greater than 20$\deg$. 
In the one-prong and three-prong hadronic decay channels
we require the total momentum of the decay products to be
greater than 11 \GeV\, and the angle between 
the direction of the vector sum of the decay products and
either jet axis to be greater than 20$\deg$.  
In addition we require the two hadronic jets to be back--to--back
within 65$\deg$.
These requirements result in a correct tau decay product identification
of 97\%, 96\%, 90\% and 89\% in the electron, muon, one-prong hadronic
and three-prong hadronic decay channels respectively,
where the tau decay products are said to be correct 
if they lie within 10$\deg$ of the original tau flight direction
and have the correct charge.
The overall selection efficiency drops from 61\% to 53\%
with respect to \cite{mass172-analysis}. 
The expected signal and
the total number of events selected from the data 
are shown in tables~\ref{tab-summary} and \ref{tabtgc-eff-back}.

\subsection{\bm Backgrounds } 
\label{sec-effback}

The estimated background contaminations from different sources are 
given in table~\ref{tabtgc-eff-back}. 
The main background process, \ZGqq, is simulated using PYTHIA. 
Most of the processes leading to four fermions in the final state 
are evaluated
using  grc4f~\cite{GRC4F} and EXCALIBUR~\cite{EXCALIBUR}.
Backgrounds from two-photon processes are evaluated using
PYTHIA. 

The background labelled in table~\ref{tabtgc-eff-back} as `4-fermion' 
represents the contribution due to  all four fermion diagrams 
excluding the three main \WW\, production diagrams 
(known as ``CC03''~\cite{LEP2YR}). 
This contribution includes all diagrams which lead to the same final 
state as the signal,
and which may therefore interfere to give a negative contribution, 
as well as those leading to different final 
states but which can be misidentified as signal. 
The 4-fermion contamination is obtained using grc4f, by subtracting 
the accepted cross section for events generated
using only the CC03 diagrams from that found for events generated using all
diagrams. 
The results were found to be in good agreement with a similar study 
using EXCALIBUR.

\begin{table}[htbp]
 \begin{center}
 \begin{tabular}{||l||r|r|r|r||r||} \hline \hline
  Channel  & Expected Signal  & \Qqln\, Bkg. & Other Bkg. & Total & Observed \\ 
\hline \hline
\multicolumn{6}{||c||}{From~\cite{mass172-analysis}
[see note in caption].} \\ 
\hline 
  \WWqqen  & $16.2 \pm 0.4 $ & $ 0.6\pm0.0 $  & $ 1.2\pm0.4 $  & $18.0 \pm0.6$ & 19   \\
  \WWqqmn  & $16.5 \pm 0.4 $ & $ 0.9\pm0.0 $  & $ 0.6\pm0.1 $  & $17.9 \pm0.4$ & 16   \\
  \WWqqtn  & $11.6\pm 0.4  $ & $ 1.9\pm0.1 $  & $ 2.7\pm0.7 $  & $16.2 \pm0.8$ & 20   \\
  \WWqqqq  & $46.8 \pm 1.2 $ &    $ - $       & $14.3\pm2.9 $  & $61.0 \pm3.1$ & 57   \\
  \WWlnln  & $10.7 \pm 0.3 $ &    $ - $       & $ 0.8\pm0.2 $  & $11.5 \pm0.4$ &  8   \\
\hline \hline
\multicolumn{6}{||c||}{After additional requirements for angular distribution analysis.}\\
\hline 
  \WWqqen  & $16.0 \pm 0.4 $ & $ 0.5\pm0.0 $  & $ 0.6\pm0.3 $  & $17.1 \pm0.5$ & 17   \\
  \WWqqmn  & $16.4 \pm 0.4 $ & $ 0.6\pm0.0 $  & $ 0.2\pm0.3 $  & $17.2 \pm0.5$ & 15   \\
  \WWqqtn  & $9.8 \pm 0.4 $  & $ 0.8\pm0.1 $  & $ 1.6\pm0.3 $  & $12.2 \pm0.5$ & 16   \\
\hline \hline
 \end{tabular}
 \end{center}
\caption{Observed number of candidate events in each \WW\, decay
  channel,
  together with expected numbers of signal and background events
  predicted by the \SM.
  The predicted numbers for
  signal include systematic uncertainties from the efficiency,
  luminosity, beam energy, \WW\, cross-section 
  and \Mw, while the background
  estimates include selection and luminosity uncertainties.
  {(Note: In this table, contamination in the \Qqln\, channels from
  other \Qqln\, channels is listed separately, whereas
  in \protect\cite{mass172-analysis} it is included in the signal.) }
}
\label{tab-summary}
\end{table}

\begin{table}[htbp]
\begin{center}
\begin{tabular}{||l||c|c|c||}\hline \hline
                 & \multicolumn{3}{|c||}{Channel }   \\
\cline{2-4}
 Source          &    \Qqen     &    \Qqmn     &  \Qqtn         \\
\hline \hline
Expected signal (fb) & $1543\pm16$ & $1579\pm13$  &  $962\pm19$   \\
\hline \hline
Background (fb)  &              &              &             \\
\WWqqen          &  $-$         & $5\pm1$      & $37\pm3$    \\
\WWqqmn          &  $0.8\pm0.5$ & $-$          & $44\pm4$    \\
\WWqqtn          &  $49\pm4$    & $59\pm4$     & $-$         \\
\WWqqqq          &  $0\pm0$     & $0.6\pm0.4$  & $8\pm2$     \\
\WWlnln          &$0.8\pm0.5$   & $0.3\pm0.3$  & $0.6\pm0.4$ \\
\ZGqq            &  $22\pm6$    & $9\pm4$      & $43\pm8$    \\
4-fermion (non-qqee) &  $-1\pm25$ & $15\pm25$    & $59\pm20$ \\
4-fermion (qqee) &  $20\pm11$   & $ 0\pm0$     & $44\pm20$   \\
Two-photon       &  $22\pm13$   & $0\pm0$      & $0\pm0$     \\
\hline      
Total  (fb)      &  $113\pm31$ & $88\pm25$  & $236\pm30$     \\ 
Total  (\%)      &  $7\pm2$    & $6\pm2$    & $25\pm3$       \\ 
\hline \hline
\end{tabular}
\end{center}
\caption{
Signal and background cross sections for the 
$\Qqen$, $\Qqmn$ and $\Qqtn$ samples used in the analysis of
angular distributions.
The errors on the expected signal include both
statistical and systematic errors. The errors on the background
estimations are statistical only.
}
\label{tabtgc-eff-back}
\end{table}

\section{Kinematic variables for the \boldmath\Qqln\, event sample }
\label{sec-data-distributions}

For each $\Qqln$ event we measure three angles
 \footnote{The definition of the axes is such that $z$ is along
 the parent W flight direction and $y$ is in the direction 
 $\vec{e^-} \times \vec{W}$
 where $\vec{e^-}$ is the electron beam direction and
 $\vec{W}$ is the parent W flight direction.  The axes are defined in the
 W rest frame.} 
 ~\cite{LEP2YR,BILENKY,SEKULIN}:
\begin{itemize}


\item[1)]  $\Cthw$, the production angle of the $\Wm$ with respect to 
the e$^-$ beam direction,

\item[2)] $\Cthstl$, the polar angle of the charged lepton 
with respect to the parent W flight direction measured in the W rest frame,

\item[3)]  $\Phistl$, the azimuthal angle of the charged lepton 
around the axis 
given by the parent W flight direction, measured in the W rest frame.



\end{itemize}

In the case of the \Qqen\, and \Qqmn\, events 
we perform a three constraint kinematic fit,
which demands energy and momentum conservation assuming a zero mass 
for the missing neutrino.
The fit minimises a $\chi^2$ sum which also includes a contribution 
from the difference between 
 \footnote{ 
 The W mass is treated as Gaussian in the kinematic fit.  However, in 
 order to simulate the expected Breit-Wigner form of the W mass spectrum,
 the variance of the Gaussian is updated at each iteration of the 
 kinematic fit in such a way that the probabilities of observing 
 the current fitted W mass are equal whether calculated
 using the Gaussian distribution or using a simple Breit-Wigner.
 }
the nominal world average W mass of 80.33 \GeV\, and  
the fitted masses of both the hadronic and leptonic systems. 
We demand that the kinematic fit converges with a probability 
of $> 10^{-3}$. For the small number of approximately 5\% of
events which fail at this point we revert to using the results of a
fit without the W mass constraints.
We then obtain $\Cthw$ by adding together
the kinematically fitted four-momenta of the two jets. 
The charges of the W bosons are determined by the 
sign of the charged lepton. 
The decay angles are obtained from the fitted charged lepton
four-momentum after boosting back to the parent W rest frame
using the fitted W four-momentum to determine the boost.


In the case of the \Qqtn\, events
we obtain $\Cthw$ by adding together
the measured four-momenta of the two reconstructed jets, and 
the charges of the W bosons are determined by the sign of the sum of the
charges of the tracks of the tau decay products.
In order to reconstruct the decay angles, the flight direction of the 
tau is approximated by the direction of its charged decay products.
The four unknown quantities can then be calculated 
using energy and momentum conservation. 
These are the energy of the tau and the three-momentum of the 
tau neutrino
originating directly from the W decay.
The decay angles are then obtained as for the other $\Qqln$ events.

\begin{figure}[htbp]
\epsfxsize \the\textwidth
\epsffile{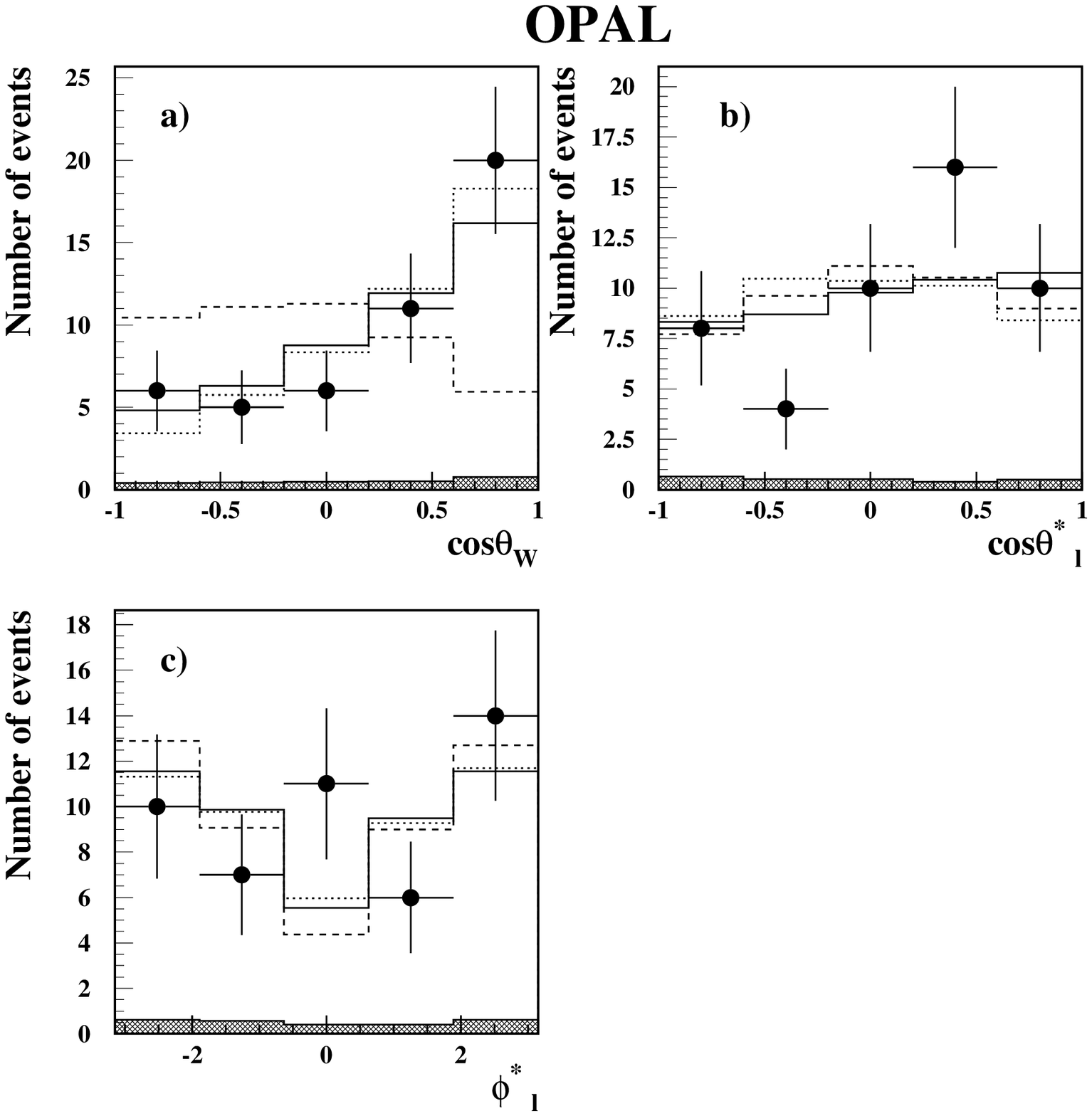}
\caption{ The distributions of the kinematic variables 
$\Cthw$, $\Cthstl$, $\Phistl$ obtained from the $\Qqln$ events. 
The hatched histogram shows the non-$\Qqln$ background.
These are compared with the distribution
expected in the \SM\, using fully simulated \MC\, events.
The predicted distributions for $\awf= +2 (-2)$ are also shown as
dotted (dashed) lines.
{\small Note: in the case of W$^+\ra\bar{\ell}\nu$ 
events the value of $\Phistl$
is shifted by $\pi$ in order to overlay W$^+$ and W$^-$ 
distributions in the same plot. }
}
\label{figtgc-qqln-distributions}
\end{figure}

In figure~\ref{figtgc-qqln-distributions} we show the distributions
of all three angles obtained from the combined 
$\Qqen$, $\Qqmn$ and $\Qqtn$ 
event sample, and the expected distributions for $\awf = \pm 2$ and 0.
The shapes of the the distributions are obtained from fully simulated
Monte Carlo event samples. The distributions are normalised to 
number of events observed in the data.

\section{Anomalous coupling analysis}

In this section we present the analysis of the \WW\, event sample in order to 
place limits upon anomalous coupling parameters.
The analysis uses the event sample in two distinct ways. 
The first part uses only the \Qqln\, events to 
constrain \tgc s through the predicted variation of
the differential distribution of the production and decay angles.
We do not distinguish between the \Qqen, \Qqmn\, and \Qqtn\, events.
The second part uses all channels to
constrain \tgc s through the variation of the expected cross-section
as a function of \tgc\, parameter. 
A log likelihood, $\LL$, curve is calculated for each part of the analysis. 
These are independent and are added together to obtain a combined
$\LL$ curve from which the results are obtained. In the following
the symbol $\alpha$ stands for a generic anomalous coupling parameter.

A binned likelihood method is used to analyse the 
three dimensional differential cross section.
We divide $\Cthw$ into 20 bins in the range [$-$1,1], 
$\Cthstl$ into 10 bins in the range [$-$1,1] and 
$\Phistl$ into 5 bins in the range [$-\pi,\pi$].
The $\LL$ distribution is obtained
in several steps.

In the first step we parameterise in each bin the expected cross-section
due to \WW\, production, 
before detector and acceptance effects, as a function of $\alpha$. 
We denote this as $\sigma_i^{\mathrm{gen}}(\alpha)$,
where $i$ refers to each bin in the three dimensional space.
It is obtained using large samples of EXCALIBUR \MC\, events
including the effects of the W width, $\Gw$, and initial state 
radiation, ISR, 
but without detector simulation. 
In each bin, $\sigma_i^{\mathrm{gen}}(\alpha)$ is parametrised
using the fact that $\sigma$ is a quadratic function of $\alpha$ (as
$\alpha$ enters linearly in the Lagrangian).
Several samples are generated for different values of $\alpha$
and a quadratic parameterisation obtained to predict
$\sigma(\alpha)$ for all other values of $\alpha$.

In the second step we calculate a correction matrix to include  the
effects of acceptance, resolution and feedthrough from other
\WW\, decay channels.  
We use fully simulated \MC\, events generated with
$\alpha=0$ to obtain factors, $c_{ki}$, which allow
for events generated in true bin $i$ being reconstructed in bin $k$.
All \WW\, decays which are reconstructed as $\Qqln$ events 
in bin $k$ are counted in the correction factors $c_{ki}$. 
The $c_{ki}$ therefore include the effect
of feedthrough from other \WW\, decay channels 
which is always a fixed fraction of the signal 
which is independent of the value of $\alpha$.  
For a given bin, corresponding to a limited phase space region,
the $c_{ki}$ factors are assumed to be independent of $\alpha$. 
Possible biases due to this assumption are discussed in section~\ref{sec-syst}.

Combining these terms results in the
expected observed cross-section for each bin $k$, due to all \WW\, channels:
\[
\sigma_k^{\small \mathrm{WW}}(\alpha)= 
\sum_{i} c_{ki}\sigma_i^{\mathrm{gen}}(\alpha)
\]

\begin{figure}[htbp]
\epsfxsize \the\textwidth
\epsfysize 19cm
\epsffile{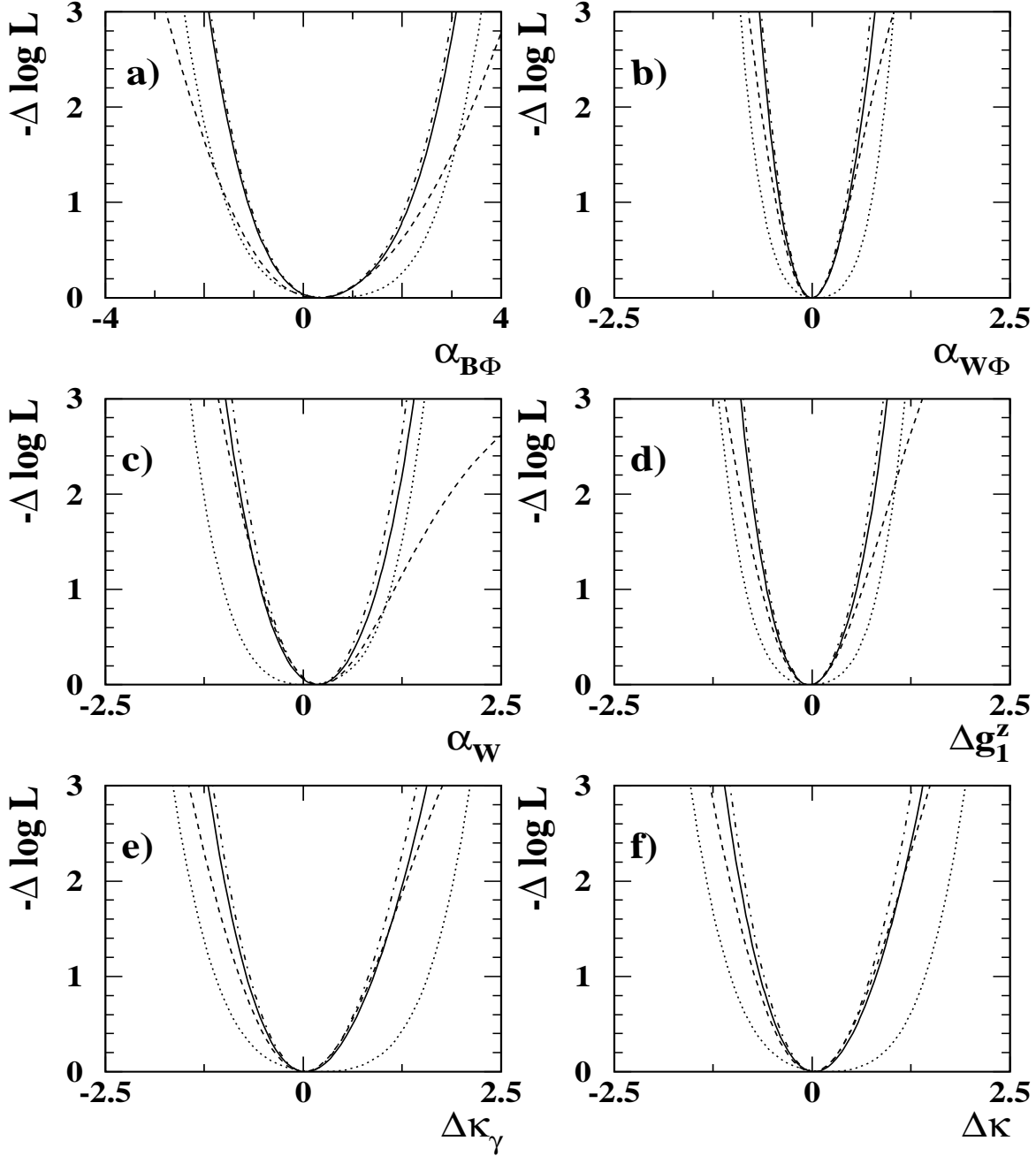}
\caption{ 
The negative log likelihood curves 
(statistical contribution only) obtained from 
the total event rate (dotted) and the shape of the 
differential distributions (dashed).  
The dot-dashed curve is obtained by adding these together.
The solid curve shows the final result obtained when
the systematic error is added.
The  $\Delta$ prefix indicates that the 
likelihood is shown relative to its minimum value.
The figures correspond to
a) $\abf$, 
b) $\awf$, 
c) $\aw$, 
d) $\dgz$, 
e) $\dkg$ assuming the HISZ constraints and
f) $\dk$ assuming $\dkg=\dkz$.
}
\label{figtgc-ll}
\end{figure}

In the next step the cross-section for the $\ZGqq$ and two photon 
backgrounds, $\sigma_k^{\mathrm{bkg}}$, is 
estimated using PYTHIA
and is added to give $\sigma_k^{\mathrm{all}}(\alpha)$.
The small background due to four fermion diagrams given in 
table~\ref{tabtgc-eff-back} is neglected. The effect of this neglect is 
discussed in section~\ref{sec-syst}.
The quantity 
$\sigma_k^{\mathrm{all}}(\alpha)$ now contains all of the information
on the shape of the expected distribution
as a function of $\alpha$. We multiply this by a scale factor to give
the prediction for the number of events in each bin, $n_k$, where
the scale factor is chosen such that the predicted total number of events
is equal to the actual number of events observed in the data.  
This is to ensure that we do not incorporate any 
information from the overall production rate
in this part of the analysis.
The probability for observing the number of events
seen in each bin for an expectation of $n_k$ 
is calculated using Poisson statistics.
The  statistical fluctuations in $c_{ki}$ and $\sigma_k^{\mathrm{bkg}}$ 
are taken into account
using the method of reference~\cite{BARLOW}.
The negative $\LL$ distributions for each of the $\alpha$ parameters
studied are shown in figure~\ref{figtgc-ll}.

The information given by the observed event rates 
is included by calculating the likelihood for the mean number
of events expected as a function of $\alpha$ to 
have resulted in the observed number of events given
in the upper part of table \ref{tab-summary}. 
This is calculated separately for each channel. 
The variation of the mean number of signal events expected as
a function of $\alpha$ is determined using fully simulated
EXCALIBUR events subjected to the same selection requirements
as the data. The number of selected events is parameterised
using the quadratic dependence mentioned earlier. The overall 
normalisation is adjusted to agree with the GENTLE prediction
at the \SM\, point.
The probability of finding the observed number of events is calculated
assuming a Poisson distribution for the signal and background in each 
channel. 
The probabilities obtained for each channel are multiplied
together and the negative $\LL$ distribution obtained. 
These are shown in figure
\ref{figtgc-ll} for each parameter considered. 

The curves due to the total and differential
cross-section information are consistent with each other 
for each of the $\alpha$ parameters. Both curves provide a useful 
constraint.  These are independent measurements and are 
added together to give the overall $\LL$ distributions shown as the 
dash-dotted lines in figure \ref{figtgc-ll}. 

\begin{figure}[htbp]
\epsfxsize \the\textwidth
\epsfysize 19cm
\epsffile{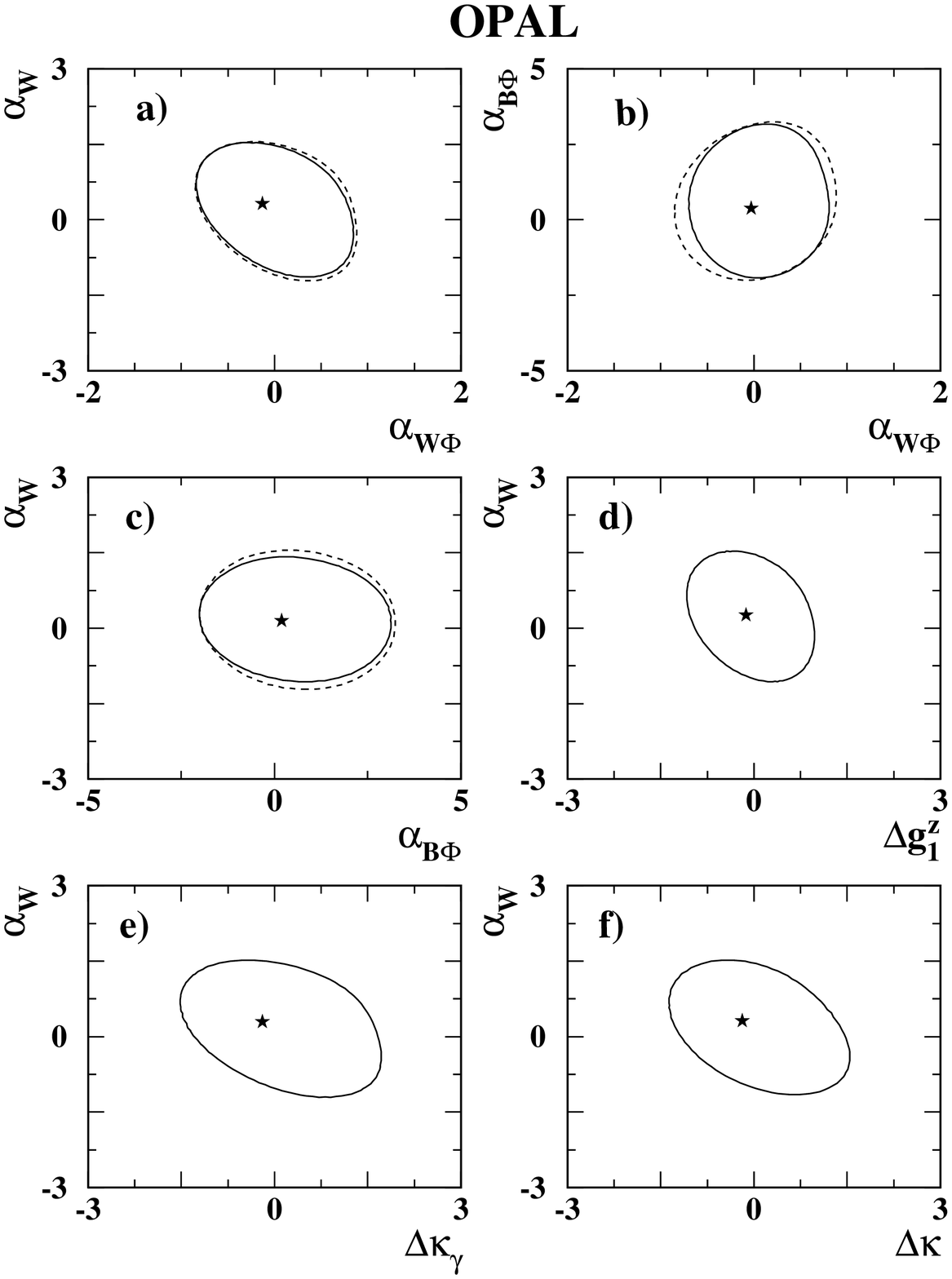}
\caption{ 
The 95\% C.L. two dimensional correlation contours 
for different \tgc\, parameters.
The effect of statistical and systematic errors is included.
The star indicates the minimum point.
The first three plots show all pairs of the $\abf$, $\awf$, and $\aw$
parameters. 
Figure d) shows $\dgz$-vs-$\aw$,
figure e) shows $\dkg$-vs-$\aw$ assuming the HISZ constraints 
and figure f) shows $\dk$-vs-$\aw$ assuming $\dkg=\dkz$.
In figures a), b) and c) the dashed line shows the 
results obtained by varying all three $\alpha$ parameters 
(see text), where both statistical and
systematic errors are included.
}
\label{figtgc-2d-contour}
\end{figure}

We also calculate the two dimensional likelihood distributions resulting 
when two anomalous couplings are simultaneously
allowed to vary from the \SM. This is incorporated  
into both parts of the analysis described above by parameterising
the predicted variation of signal events using a 
quadratic function of both \tgc s. 
The resulting 95\% probability contours for different pairs of couplings 
are shown in figure \ref{figtgc-2d-contour}.
We also perform the analysis allowing all three of the $\alpha$
parameters to vary simultaneously. 
The resulting set of two dimensional correlation contours are also 
shown in figure \ref{figtgc-2d-contour} a), b) and c),
where in each case the third parameter 
is varied in order to re-minimise the $\LL$ at each point
in the two dimensional space. 

\vspace{5mm}
{\bf Alternative analyses }
\vspace{1mm}

We have developed two other independent analyses, which are largely
complementary.
The first method uses a simple unbinned maximum likelihood
approach which does not include the effects of $\Gw$, ISR,
acceptance, resolution or background in the analysis of the kinematic
variables. These effects are expected to be small in comparison
to the present level of statistical precision, and so the method
provides a robust systematic comparison. 
The second method uses an optimal observables 
method~\cite{DIEHL,PAPADOPOO} which includes all detector
effects. 
The $\LL$ curves produced by all three methods agree very well
and give us confidence in the stability of the results.

\section{Systematic error studies}
\label{sec-syst}

The \MC\, simulation of the measured quantities depends
mainly upon the simulation of the jets from the W hadronic decay.
Jet reconstruction has been studied and tuned extensively at \LepI, 
showing good agreement between distributions measured from data and 
\MC\, samples.
We therefore expect this 
to be adequate for the small number of events in this sample. 
Studies of back-to-back jet pairs using the \LepI\, data 
yield the following possible differences between data and
the \MC\, simulation of jets;
10\% for energy resolution, 
0.5\% for energy scale and
10\% for resolution in $\cos\theta$ and $\phi$.
As a direct test of $\Cthw$ reconstruction
we have used radiative $\Zqq$ events taken from the 
$\sqrt{s}=91$, 161 and 172~\GeV\, data. 
By selecting events containing 
observed radiated photons with energies up to 20 \GeV\, we
obtain a sample of jet pairs exhibiting a similar
acollinearity distribution to W decays. 
Assuming that no other photons have been radiated in the event
the true direction of the $\Zqq$ system is 
opposite to that of the photon and the difference between the
value of $\cos\theta$ measured from the photon and that
measured from the hadronic system is therefore
strongly related to the resolution of $\Cthw$. 
The results obtained from both data
and \MC\, events agree well showing no significant
differences in shape or width of the distributions. 
We obtain a conservative upper limit upon
the possible relative shift on $\Cthw$ between data and \MC\, of 0.01.
All of these variations are then used to vary the \MC\, jet
reconstruction in the analysis and 
the resulting changes caused to the \tgc s were added in quadrature 
and taken as a systematic error. These are shown in table \ref{tabtgc-syst}
labelled as a).


We have evaluated the systematic errors due to several other factors.
\begin{itemize} 
\item[b)]  The effect of limited \MC\, statistics. 
\item[c)]  The uncertainty in the world average measured 
W mass and the LEP beam energy.
\item[d)]  The uncertainties of the  overall normalisation
(selection efficiency and luminosity)
given in~\cite{mass172-analysis} used in the calculation of the $\LL$ curve 
derived from the event rate.
\item[e)]  We have varied the \MC\, generator to use both  EXCALIBUR and 
KORALW~\cite{KORALW} to generate the \SM\, sample. 
We have varied the fragmentation model by using HERWIG~\cite{HERWIG}.
\item[f)] We have varied the backgrounds within the estimated systematic
errors given in~\cite{mass172-analysis}  
in the calculation of the $\LL$ curve derived from the observed event rate.
We have doubled and removed the 
\Zgamma\, and two-photon background additions used
in the calculation of the $\LL$ curve derived from the differential 
distributions. 
\item[g)] The analysis was carried out using \MC\, samples generated
at a nominal energy of 171 \GeV. We have estimated the possible
bias this introduces by using \MC\, samples generated at 172 \GeV\
as test data.
\end{itemize}
The errors due to these sources are listed in table~\ref{tabtgc-syst}. 

This analysis relies upon EXCALIBUR correctly describing the variation of
cross section, angular distribution and four-fermion background 
as functions of anomalous \tgc s. 
We have investigated the limits of this assumption by comparing these 
quantities to those predicted by several other programs
(GENTLE, KORALW and ERATO~\cite{ERATO}). 
We find that the total cross section predictions agree to within 
approximately 2.5\% across a wide range of anomalous \tgc\, values
and we add this variation as a further systematic error to d).
We find that both the angular distributions and the contributions
of diagrams in addition to the three \WW\, diagrams are also
compatible between the programs at the level of a few percent
and we assign no extra systematic error.

Finally we consider possible biases of the method.
This method allows for all effects of $\Gw$, ISR, acceptance and detector
resolution in the calculation of both $\LL$ curves and is
therefore in principle free from biases due to these effects. 
However biases may be introduced by the coarse binning, by the 
approximation in the correction used to include detector effects,
or by the neglect of the small four fermion background.
To quantify these effects  we perform the analysis
using samples of approximately 20,000 fully simulated 
four fermion \MC\, events
produced by both grc4f and EXCALIBUR as input. 
These samples were generated
with different TGC values.
In all cases the \tgc\, value reconstructed by the analysis 
is found to be consistent with the generated value and no
significant difference was observed between the grc4f
and EXCALIBUR samples.
The greater of the statistical precision of the test or the
measured bias was taken as 
a conservative systematic error and is shown in row h) of 
table~\ref{tabtgc-syst}. 

All contributions to the systematic errors are summarised in table
\ref{tabtgc-syst}. 
To include the systematic error in the likelihood curves, we apply
a simple procedure which would be exact if 
all errors were Gaussian.
We calculate a function of $\alpha$, 
$\Delta\LL(\alpha)_{syst}$, 
to add to the $\LL$ curves. This is given by
\[
\Delta\LL(\alpha)_{syst} =
(\alpha - \alpha_0)^T 
[ (V_{stat} + V_{syst})^{-1} - (V_{stat})^{-1} ]
(\alpha - \alpha_0)
\]
where $\alpha$ represents a vector of one or more \tgc\, parameters,
$\alpha_0$ is the \tgc\, value at the minimum of the $\LL$ curve,
$V_{syst}$ is the systematic error matrix constructed
assuming the errors are Gaussian, 
and $V_{stat}$ is approximated (for this purpose
only) by assuming a parabolic $\LL$ distribution around $\alpha_0$.
The full $\LL$ curves for each parameter are obtained by adding
$\Delta\LL(\alpha)_{syst}$ to the $\LL$ curve due to statistical errors only,
and these are shown as solid lines in figures \ref{figtgc-ll}
and \ref{figtgc-2d-contour}

\begin{table}[htbp]
\begin{center}
\begin{tabular}{||l|c||c|c|c|c|c|c||}\hline \hline
\multicolumn{2}{||c||}{Source}&  \multicolumn{6}{c||}{Error on parameter}  \\
\multicolumn{2}{||c||}{ } & $\abf$ & $\awf$ & $\aw$ & $\dgz$ & $\dkg$ & $\dk$\\
\hline \hline                               
a) &Jet reconstruction    &  0.21  &  0.05  &  0.03 &  0.05 &  0.12 & 0.12 \\
b) &MC statistics         &  0.06  &  0.04  &  0.05 &  0.05 &  0.06 & 0.07 \\
c) &W mass/LEP energy     &  0.04  &  0.01  &  0.01 &  0.00 &  0.01 & 0.01 \\
d) &Overall normalisation &  0.07  &  0.01  &  0.02 &  0.01 &  0.02 & 0.02 \\
e) &MC generator          &  0.11  &  0.02  &  0.15 &  0.02 &  0.04 & 0.08 \\
f) &Background            &  0.22  &  0.06  &  0.12 &  0.06 &  0.11 & 0.09 \\
g) &171/172 \GeV\, diff.   &  0.12  &  0.05  &  0.10 &  0.08 &  0.07 & 0.07 \\
h) &Bias tests            &  0.11  &  0.04  &  0.05 &  0.05 &  0.06 & 0.06 \\
\hline                                                  
   &Total                 &  0.38  &  0.11  &  0.23 &  0.14  &  0.20 &  0.21 \\
\hline \hline
\end{tabular}
\end{center}
\caption{ Contributions to the systematic errors in the determination of 
\tgc\, parameters.}
\label{tabtgc-syst}
\end{table}

\section{Results and Conclusion}

We express the results as measurements with the one
standard deviation limits given by the values of $\alpha$ where 
the negative $\LL$ curve rises by 0.5 from the minimum. 
The results are given in table \ref{tabtgc-results} where 
the statistical and systematic errors are shown separately. 
The expected statistical errors for a sample of this size were evaluated 
by repeating the analysis using \MC\
events divided into sub-samples equivalent to the actual 
data sample. The RMS spread and the average error agreed within 10\%,
and the latter is shown in the table.

We also express the results as  95\% confidence level (C.L.) limits
given by the values of $\alpha$ where 
the change in likelihood is 1.92. The ranges obtained 
are shown in table \ref{tabtgc-results}.


In our previous publication~\cite{tgc161-analysis} we have presented a 
measurement of the $\awf$ parameter using the 161 \GeV\, data. 
By including the $\LL$ obtained from that measurement  
in this analysis we obtain the combined result 
\awf=\resawfc .
The corresponding 95\% C.L. range is
$-0.59 < \awf < 0.52$.

The results which we have obtained for the parameters  
listed as c), e) and f) in the results table 
can be compared with those reported
recently by CDF~\cite{CDF} and D0~\cite{D0}
which are a factor of between two and three more precise. The measurements
agree in all cases.


\begin{table}[htbp]
\begin{center}
\begin{tabular}{||l||c|c||c||} \hline \hline
Parameter & Measurement & Expected   & 95\% C.L. range   \\
          &             & stat. err. &                   \\ \hline
          &             &            &                   \\
\abf      & \resabf     &   $\pm1.1$ & [$-1.6 ,~~2.7 $]  \\
          &             &            &                   \\
\awf      & \resawf     &   $\pm0.29$& [$-0.55,~~0.64$]  \\
          &             &            &                   \\
\aw       & \resaw      &   $\pm0.48$& [$-0.78,~~1.19$]  \\
          &             &            &                   \\
\dgz      & \resdgz     &   $\pm0.39$& [$-0.75,~~0.77$]  \\
          &             &            &                   \\
\dkH      & \resdkH     &   $\pm0.53$& [$-0.98,~~1.24$]  \\
          &             &            &                   \\
\dkg=\dkz & \resdk      &   $\pm0.50$& [$-0.90,~~1.12$]  \\
          &             &            &                   \\
\hline
\hline
 \end{tabular}
 \end{center}
 \caption{ 
    Measurements of anomalous coupling parameters 
    using the OPAL 172 \GeV\, data sample.
    The measured values and one standard deviation errors
    are shown in column 2, where the statistical (asymmetric) and systematic 
    (symmetric) errors are shown separately. 
    The expected error obtained from multiple \MC\, samples is 
    shown in the third column.
    The 95\% confidence level ranges including systematic
    errors are shown in the fourth column.
  }
 \label{tabtgc-results}
\end{table}

In conclusion, 
we have presented results using \WW\, events 
reconstructed from \intLdt\Ipb\, of data recorded at \LepII\, at a 
centre-of-mass energy of 172 \GeV. 
We have used the observed \WW\, event rate
and the distribution of kinematic variables
to place limits upon possible
anomalous triple-gauge-boson couplings. 
We measure
\abf=\resabf ,
\awf=\resawf ,
\aw=\resaw ,
\dgz=\resdgz ,
\dkH=\resdkH , and
\dk=\resdk .
These results agree with the \SM\, expectation of zero for each 
anomalous coupling. 
The negative $\LL$ distributions for these parameters
are shown in figure~\ref{figtgc-ll}.
The two dimensional correlation contours are
shown in figure~\ref{figtgc-2d-contour}.

\section*{Acknowledgements}
\par
We particularly wish to thank the SL Division for the efficient operation
of the LEP accelerator at all energies
 and for
their continuing close cooperation with
our experimental group.  We thank our colleagues from CEA, DAPNIA/SPP,
CE-Saclay for their efforts over the years on the time-of-flight and trigger
systems which we continue to use.  In addition to the support staff at our own
institutions we are pleased to acknowledge the  \\
Department of Energy, USA, \\
National Science Foundation, USA, \\
Particle Physics and Astronomy Research Council, UK, \\
Natural Sciences and Engineering Research Council, Canada, \\
Israel Science Foundation, administered by the Israel
Academy of Science and Humanities, \\
Minerva Gesellschaft, \\
Benoziyo Center for High Energy Physics,\\
Japanese Ministry of Education, Science and Culture (the
Monbusho) and a grant under the Monbusho International
Science Research Program,\\
German Israeli Bi-national Science Foundation (GIF), \\
Bundesministerium f\"ur Bildung, Wissenschaft,
Forschung und Technologie, Germany, \\
National Research Council of Canada, \\
Hungarian Foundation for Scientific Research, OTKA T-016660, 
T023793 and OTKA F-023259.\\




\newpage

\end{document}